\documentclass[aps,pra,superscriptaddress,twocolumn,showpacs,titlepage=false]{revtex4-1}
\usepackage{graphicx}
\usepackage{amsmath}
\usepackage{amssymb}
\usepackage{subfigure}
\usepackage{natbib}
\usepackage{color}
\usepackage[breaklinks=true,colorlinks=true,linkcolor=blue,urlcolor=blue,citecolor=blue]{hyperref}
\def\bra#1{{\left\langle #1 \right|}}
\def\ket#1{{\left| #1 \right\rangle}}

\usepackage{xcolor}	 
\definecolor{jesusgreen}{RGB}{0,127,0}

\usepackage[T1]{fontenc}
\usepackage{tabularx}
\usepackage{graphicx}
\usepackage{amsmath}
\usepackage{amsfonts}
\usepackage{verbatim}
\usepackage{bbm, bm}
\usepackage{amssymb}
\usepackage{multirow}
\usepackage{enumitem}

\usepackage{calc}
\usepackage{graphicx}
\usepackage{amsmath}
\usepackage{amssymb}
\usepackage{dsfont}
\usepackage{relsize}
\usepackage{multirow}
\usepackage{rotating}
\usepackage{bm}
\usepackage{url}
\usepackage[thinlines]{easytable}
\usepackage{blkarray}

\begin{document}

\title{A hybrid machine-learning algorithm for designing quantum experiments}
\author{L. O'Driscoll}
\affiliation{Centre for the Mathematics and Theoretical Physics of Quantum Non-Equilibrium Systems (CQNE), School of Mathematical Sciences, University of Nottingham, University Park, Nottingham, NG7 2RD, UK.}
\author{R. Nichols}
\affiliation{Centre for the Mathematics and Theoretical Physics of Quantum Non-Equilibrium Systems (CQNE), School of Mathematical Sciences, University of Nottingham, University Park, Nottingham, NG7 2RD, UK.}
\author{P. A. Knott}
\thanks{Contact email address: Paul.Knott@nottingham.ac.uk}
\affiliation{Centre for the Mathematics and Theoretical Physics of Quantum Non-Equilibrium Systems (CQNE), School of Mathematical Sciences, University of Nottingham, University Park, Nottingham, NG7 2RD, UK.}
\date{\today}

\begin{abstract}
We introduce a hybrid machine-learning algorithm for designing quantum optics experiments to produce specific quantum states. Our algorithm successfully found experimental schemes to produce all $5$ states we asked it to, including Schr\"odinger cat states and cubic phase states, all to a fidelity of over $96\%$. Here we specifically focus on designing realistic experiments, and hence all of the algorithm's designs only contain experimental elements that are available with current technology. The core of our algorithm is a genetic algorithm that searches for optimal arrangements of the experimental elements, but to speed up the initial search we incorporate a neural network that classifies quantum states. The latter is of independent interest, as it quickly learned to accurately classify quantum states given their photon-number distributions.
\end{abstract}
\maketitle


As artificial intelligence (AI) and machine learning develop, their range of applicability continues to grow. They are now being utilised in the fast-growing field of quantum machine learning \cite{dunjko2017machine,biamonte2017quantum,schuld2015introduction}, with one particular application demonstrating that AI is an effective tool for designing quantum physics experiments \cite{knott2016search,krenn2016automated,melnikov2018active,arrazola2018machine,sabapathy2018near}. In this vein, here we introduce a hybrid algorithm that designs and optimises quantum optics experiments for producing a range of useful quantum states, including Schr\"odinger cat states \cite{ourjoumtsev2007generation,huang2015optical,etesse2015experimental} and cubic phase states \cite{gottesman2001encoding}.

The core of our algorithm, named AdaQuantum \footnote{The algorithm, AdaQuantum, is named after Ada Lovelace, the worlds first computer programmer, and resident of Nottingham, where our own algorithm was born.} and introduced in \cite{Rosanna} and \cite{knott2016search}, uses a genetic algorithm to search for optimal arrangements of quantum optics experimental equipment. Any given arrangement will output a quantum state of light, and the algorithm's task is to optimise the arrangement to find states with specific properties. To assess the suitability of a given state, we require a \emph{fitness function} that takes as input a quantum state and outputs a number -- the \emph{fitness value} -- that quantifies whether the state has the properties we desire or not. Our previous works largely focused on quantum metrology, where our algorithm found quantum states with substantial improvements over the alternatives in the literature \cite{Rosanna,knott2016search}. While in \cite{Rosanna,knott2016search} the fitness function assessed the phase-measuring capabilities of the states, in this paper instead we look at producing a range of useful and interesting states (introduced below) to a high fidelity, and hence we use as our fitness function the fidelity to our target states.

The search space for our genetic algorithm is huge, which means that typically the algorithm has to simulate and evaluate a vast number of quantum optics experiments in order to finally find strong solutions. This introduces a major challenge in our work: the speed, efficiency, and effectiveness of our algorithm depend in a large part on simulating and evaluating the experiments as quickly and accurately as possible. If short-cuts could be found that allow a given experimental set-up to be evaluated approximately without the full simulation being performed, then this would greatly improve our optimisation: the approximate evaluations would provide a quick guide for the search to progress in the right direction, and subsequently the exact evaluation, using the full quantum simulation, would provide the precise fitness value, thus validating and fine-tuning the search. Efficient computational models of this sort for approximating the fitness function are often known as surrogates or meta-models \cite{jin2011surrogate}.

In this paper we use one such approximation during the evaluation stage: instead of explicitly calculating the fidelity of our output state to the desired states, we use a deep neural network (DNN) that learns to classify what type of state has been outputted, and approximately how close this state is to the target state. Explicitly calculating the fidelity the large number of times required to run our algorithm takes a surprisingly long time, whereas using our DNN we get a useful, albeit modest, speed-up. While in the work presented here the speed-up is only small, our results can be seen as a proof of principle, paving the way for much more demanding fitness functions, such as the Bayesian mean squared error \cite{rubio2018non,Rosanna}, to be approximately evaluated in this way. Once the DNN has guided the search in the right direction, the fidelity is then used to provide the exact fitness function (see Section \ref{sec:usingageneticalgorithmtodesignexperiments} for the full description of our algorithm).

Our DNN for classifying quantum states is likely to be of independent interest, as it quickly learnt to recognise a quantum state just given its photon-number distribution. This has potential uses in a wide range of applications, such as quantum computing or quantum cryptography, where it can be valuable to quickly recognise what type of quantum state has been produced.

Our hybrid algorithm, utilising a genetic algorithm to perform the automated search and a neural network to classify the outputted quantum states, quickly and efficiently found new quantum optics experiments to produce a range of useful quantum states to a high fidelity. This complements the recent work of Arrazola \emph{et al.} \cite{arrazola2018machine} who used machine learning to also find quantum optics experiments to produce a range of specific states, with one key difference being that our experiments are specifically designed to be practical, which comes at the cost of our states being of a lower fidelity than those in \cite{arrazola2018machine}.

This paper is structured as follows. First we introduce a general experimental scheme for engineering optical quantum states, and introduce the various quantum-optics experimental elements that are typically used. We then introduce the main goal of our work: finding experimental set-ups to create specific quantum states. We then describe the methods we use: firstly the genetic algorithm, which searches through different arrangements of the experimental equipment to find those that produce our desired states; and secondly our neural network, which classifies quantum states and enables a speed-up in our search algorithm. Finally, we present our results.

\section{Quantum state engineering with light}

\begin{figure}
\centering
\includegraphics[scale=1.55]{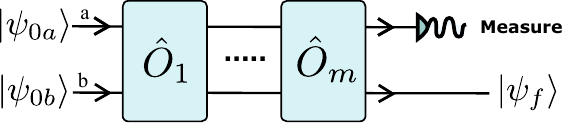}
\caption{The state engineering scheme we consider begins with two input states, $|\psi_{0a}\rangle$ and $|\psi_{0b}\rangle$, which are input into the two modes. The states then subsequently pass through a number of operators $\hat{O}_i$. To produce the final quantum state $|\psi_f\rangle$ a heralding measurement is performed on one mode.}
\label{fig:Random_general_scheme}
\end{figure}

The state-engineering scheme we use is shown in Fig.~\ref{fig:Random_general_scheme}. Firstly, two input states, $|\psi_{0a}\rangle$ and $|\psi_{0b}\rangle$, are input into the two modes. The two modes then pass through a sequence of operators $\hat{O}_i$ where $i =1,..,m$, and the final step is to perform a heralding measurement on one mode, producing the final state $|\psi_f\rangle$. With appropriate choices of input states, operators, and measurements this scheme is able to replicate a wide range of the quantum state engineering protocols in the literature, such as \cite{ourjoumtsev2007generation,huang2015optical,etesse2015experimental,bartley2012multiphoton,gerrits2010generation}. The main difference between the present paper and the usual methods is that we don't \textit{choose} the input states, operators, and measurements, but instead we use a genetic algorithm to \emph{search} for states with the desired properties (see Section \ref{sec:usingageneticalgorithmtodesignexperiments} for details of the algorithm).

Our objective is to find \emph{practical} experiments, and so we construct our state engineering protocols from elements of an experimentally-ready toolbox of quantum optics states, operators, and measurements, which is summarised in the table below.

\begin{center}
    \begin{TAB}(r, 0em, 1em)[2pt]{|c|c|}{|c|cccc|}
        {\bf Input States} & {\bf Apparatus} \\
        Fock, $\ket{n}$ & Beam splitter, $\hat{U}_{T}$ \\
        Coherent, $\ket{\alpha}$ & Phase shift, $e^{i\hat{n}\theta}$ \\
        Squeezed, $\ket{z}$ & Displacement, $\hat{D}(\alpha)$ \\
        TMSV, $\ket{z}_{12}$ & Number measurement, $\ket{n}\bra{n}$\\
    \end{TAB}
\end{center}

Here we only introduce the most important details of the toolbox; more details can be found in Appendix A. Firstly, the input states we include are the single-mode squeezed vacuum $|z\rangle$, the two-mode squeezed vacuum (TMSV) $|z\rangle_{12}$, the coherent state $|\alpha\rangle$, and Fock states $|n\rangle$; the parameters $z$, $\alpha$ and $n$ are constrained by what is possible experimentally \cite{mehmet2011squeezed,muller2015coherent,claudon2010highly,morin2012high,ourjoumtsev2006quantum,huang2015thesis}.

Next are the operators, of which the most important is the beam splitter $\hat{U}_{T}$, where $T$ is the probability of transmission, which serves to mix and entangle the two modes, enabling more exotic and useful states to be produced when part of the entangled state is measured. The other operators we use are the displacement operator $\hat{D}(\alpha)$ and the phase shift $e^{i\hat{n}\theta}$. In addition we include the identity operator $\hat{\mathbb{I}}$, because we are promoting the easiest-to-implement schemes, which would contain as many identities as possible. 

The final step of the state engineering scheme is to perform a heralding measurement on one mode of the final state. If, for example, we wish to herald on the one photon state, we can perform a photon-number resolving detection (PNRD), and only keep the output state if one photon is detected. A measurement outcome of one photon therefore \emph{heralds} the desired final state. The heralding measurement corresponds to acting on the two-mode, pre-measurement state with $\langle 1|\otimes \hat{\mathbb{I}}$, followed by normalisation. We are then left with the single mode final state $|\psi_f\rangle$. Recent progress in PNRD (for example using transition edge sensors \cite{humphreys2015tomography,gerrits2010generation}) has enabled detections of larger numbers of photons possible, so here we allow for heralding number measurements of up to $8$ photons. 

Many more states, operators and measurements can be included in this toolbox. As discussed in our paper that fully introduces our algorithm \cite{Rosanna}, the algorithm is design with flexibility in mind, so it is straightforward for more elements to be added (or removed) from the toolbox, depending on the available equipment or desired goal. But here we consider a simplified toolbox to discover whether such a limited toolbox of experimentally viable elements can still produce a range of quantum states to a high fidelity. In \cite{Rosanna} we include experimental noise, but in this work we stick to pure states and perfect operators/measurements.

\section{Finding experiments to engineer specific quantum states}
\label{eng_states}

\begin{figure}[]
	\centering
	\includegraphics[width=1\linewidth]{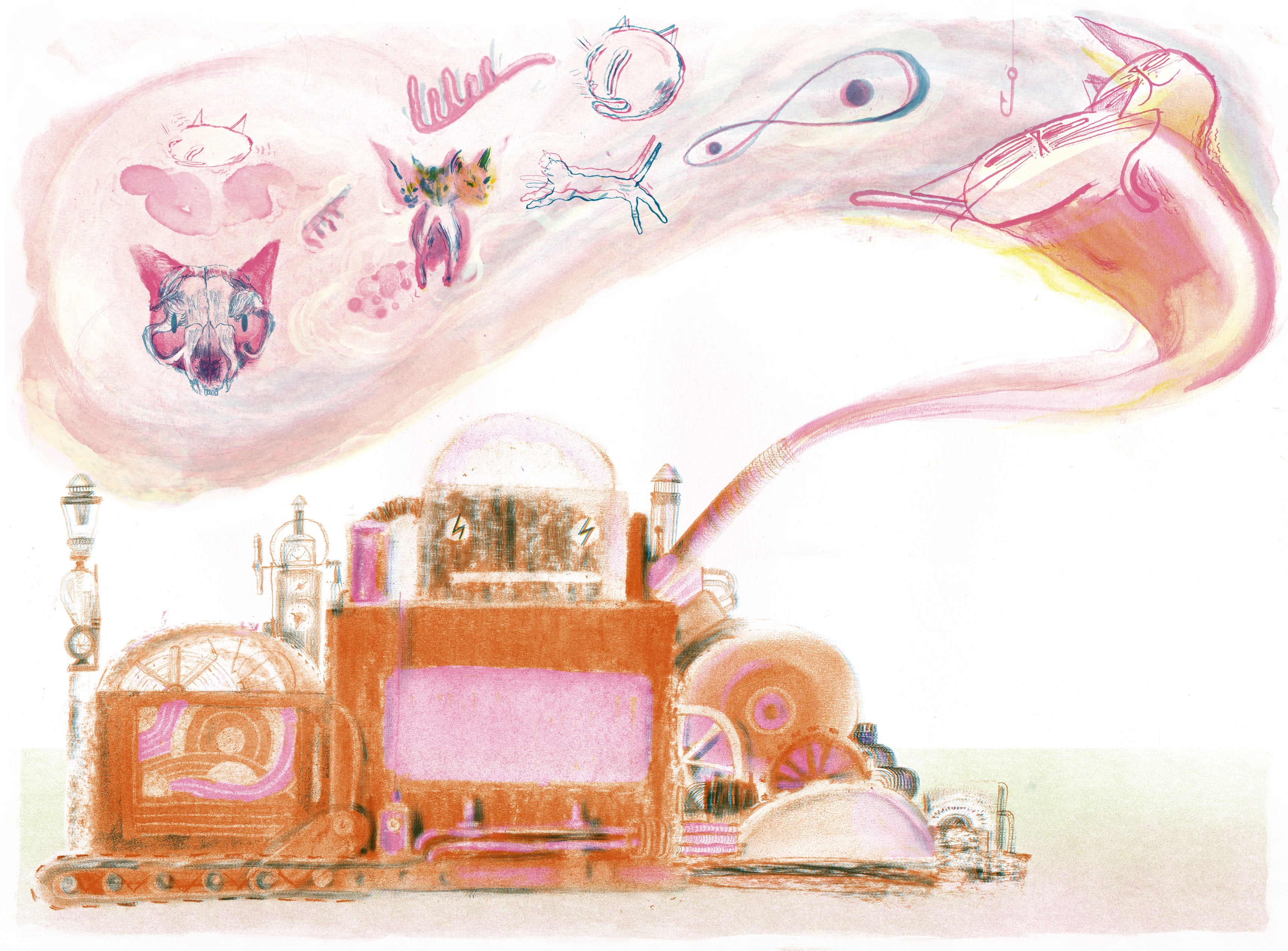}
	\caption{An artist's impression of our algorithm, AdaQuantum, which designs experiments for engineering quantum states. The algorithm is designed for flexibility, and can produce a wide range of quantum states: the illustration depicts, among others, the production of a Schr\"odinger-cat state, a three-headed cat state \cite{lee2015quantum}, and a GKP state \cite{gottesman2001encoding}. Artwork by Joseph Namara Hollis.}
	\label{fig:Ada}
\end{figure}

The task we set our algorithm, AdaQuantum, is to find experimental designs to produce a range of specific ``target'' states to a high fidelity (where the fidelity between two pure states $\ket{\psi}$ and $\ket{\phi}$ is defined as $F \equiv |\bra{\psi}\phi\rangle |^2$). The target states are shown in the table below (normalisation is omitted): these states have a range of properties and are studied and used by both theorists and experimentalists. Fig~\ref{fig:Ada} shows an artist's impression of our algorithm, AdaQuantum, producing a range of quantum states. \\

\begin{center}
    \begin{TAB}(r, 0em, 1em)[2pt]{|c|c|}{|c|c|c|c|c|c|}
        {\bf Name} & {\bf State} \\
        Cat & $\ket{\text{cat}} \backsim \ket{\alpha} + e^{i\theta}\ket{-\alpha}$ \\
        Squeezed Cat & $\hat{S}(z) \ket{\text{cat}}$ \\
        Zombie & $\ket{\alpha} + \ket{e^{2\pi i/3}\alpha} + \ket{e^{4\pi i/3}\alpha}$ \\
        ON & $\ket{0} + \delta \ket{n}$ \\
        Cubic Phase & $\exp \left (i \gamma \hat{q}^3 \right ) \hat{S}(z) \ket{0}$ \\
    \end{TAB}
\end{center}
~\\

Here $\delta,\gamma \in \mathbb{R}$; $\hat{S}(z)$ is the squeezing operator, given by $\hat{S}(z)=\exp{ \left[ {1 \over 2} (z^* \hat{a}^2 - z \hat{a}^{\dagger 2}) \right] }$, where $z \in \mathbb{C} $ and $\hat{a}^\dagger$ and $\hat{a}$ are the creation and annihilation operators, respectively \cite{barnett2002methods}; and $\hat{q}$ is the position quadrature operator, given by $\hat{q} = (\hat{a}+\hat{a}^{\dagger})/\sqrt{2}$.

The first state we search for is the optical Schr\"odinger cat state. This state is inspired by Schr\"odinger's famous thought experiment in which he proposed to put a macroscopic system -- a cat -- in a superposition of two distinct states \cite{schrodinger1935gegenwartige}. The implications of this thought experiment still spark heated debate and disagreement, but what escapes controversy is that it would be both important and interesting to create a macroscopic superposition of two distinct states. The optical Schr\"odinger cat state is moving towards this goal, as it is a superposition of two distinct coherent states, $\ket{\alpha}$ and $\ket{-\alpha}$. While the magnitude of $\alpha$ so far produced in experiments is far from making the state macroscopic, the optical Schr\"odinger cat state might eventually be produced as a macroscopic superposition, given that $\ket{\alpha}$ is the state produced from a `perfect' laser. Optical Schr\"odinger cat states (often referred to as just `cat states') have been produced in a number of experiments, such as \cite{ourjoumtsev2007generation,huang2015optical,etesse2015experimental}.

The next two states are derived from the Schr\"odinger cat state. First we can consider \emph{squeezing} a cat state by applying the squeezing operator, as shown in the table. The resulting \emph{squeezed cat state}, whilst in one sense being more exotic than the cat state, isn't necessarily more difficult to produce experimentally \cite{huang2015optical,etesse2015experimental}. The squeezed cat state can, for example, provide substantial enhancements in quantum metrology \cite{knott2016practical}. Next, instead of making a superposition of \emph{two} coherent states with different phases as in the cat state, we can make a superposition of \emph{three}, with the relative phases now differing by $2\pi /3$. Such a state can be called a three-headed cat state \cite{lee2015quantum_cat,jiang2016dynamics}, but here we prefer the name \emph{zombie cat state}, as it represents the superposition of three distinct macroscopic states: dead, alive, and undead.

We used the algorithm to search for cat states, squeezed cat states, and zombie cat states, with the following parameters: $\alpha \in [0,2]$, $\theta \in [0,2\pi]$, and $z \in \mathbb{C}$ with $|z| \in [0,1.4]$.

Next we consider the \emph{ON state}, which is a superposition of the vacuum $\ket{0}$ with an n-photon state $\ket{n}$. By controlling the relative weighting $\delta$ it has been shown that the quantum Fisher information of this state, which is often used to quantify the state's phase-measuring potential, can be made arbitrarily large whilst keeping the photon number arbitrarily small \cite{rivas2012sub} (but this state cannot actually achieve infinitely-precise measurements \cite{hall2012heisenberg,hall2012universality}). These states are also important for continuous variable quantum computation because they can be used as a resource to implement cubic phase gates \cite{sabapathy2018states}. The latter enable universal quantum computation \cite{kok2010introduction} (a non-Gaussian gate such as a cubic phase gate is essential, as Gaussian gates alone cannot be universal \cite{kok2010introduction}). The ON states we searched for have $\delta \in [0,1]$ and $n \in [1,10]$.

Another state useful for continuous variable quantum computation is the \emph{cubic phase state}, which too can be used as a resource to enable cubic phase gates \cite{gottesman2001encoding,sabapathy2018states,ghose2007non,takagi2018convex,gu2009quantum}. The cubic phase states we searched for have $\gamma \in [0,0.25]$ and $z \in \mathbb{R}$ with $z \in [0,1.4]$. Note that ``true'' cubic phase states are only obtained for infinite squeezing.

\section{Methods}
\label{Methods}

\subsection{Using a genetic algorithm to design experiments}
\label{sec:usingageneticalgorithmtodesignexperiments}





We will first introduce the general method of using a genetic algorithm (GA) to design experiments, before describing in more detail how our specific algorithm works. In order to use a GA, we must encode each possible arrangement of states, operators and measurements into a vector, which is known as a \emph{genome}. The genome contains all the information necessary to re-construct a given experimental setup, including all the parameters of the experimental elements. The GA then starts by creating a collection of genomes, which together are known as the \emph{population}. Next, the experimental setup corresponding to each genome is simulated, and the fitness function for each output state is evaluated. The fitness function must take a quantum state as an input, and output a number, the fitness value. The latter quantifies whether the states has the properties we desire or not. In this paper our fitness function will be a measure of how close the output state is to one of those states we are searching for, which are introduced in Section \ref{eng_states}. But as discussed in \cite{Rosanna}, the flexibility of our algorithm allows for a wide range of fitness functions to be used to find quantum states for any number of applications.

\begin{figure}[]
	\centering
	\includegraphics[width=1\linewidth]{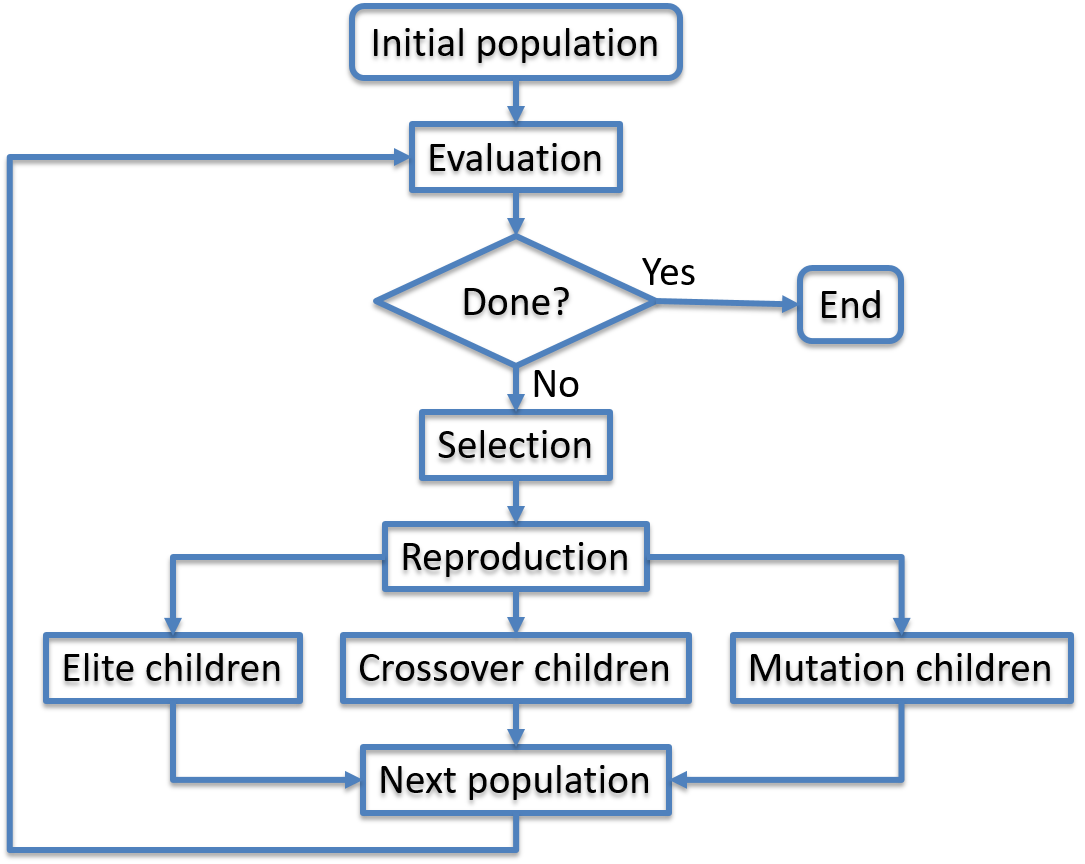}
	\caption{Flowchart of a genetic algorithm. We first create an initial \textit{population} of genomes, we evaluate these using the fitness function, and then select the `fittest' genomes. We then create a new population, named the \textit{children}, using three methods, \textit{elite}, \textit{crossover} and \textit{mutation}, which are described in the main text. The process then repeats until we have found a genome that fulfils our requirements, or until some exit criteria are met (e.g. given by the maximum allowed running time). }
	\label{fig:geneticalgorithmflowchart}
\end{figure}

After the population of genomes is assessed for their fitness, the ``fittest'' genomes -- i.e. the genomes with the largest fitness values -- are then selected, and a new population of genomes (the \emph{children}) is generated by mixing some of the genomes together (\emph{crossover}), by randomly modifying (\emph{mutating}) others, and keeping some genomes unchanged (the \emph{elite children}). This next population should, in principle, be comprised of genomes that are ``fitter'' than before. This process repeats through a number of \emph{generations}, until it is unlikely that any more generations will result in improvements. At this stage, if the algorithm has been designed appropriately, then the fittest genomes will encode optimised solutions. Through this process, our GA evolves quantum experiments that are highly suited to the task at hand. A flow chart of our algorithm is given in Fig.~\ref{fig:geneticalgorithmflowchart}.

In order to assess each genome, we must simulate the quantum optics experiment that this genome encodes, then evaluate the fitness function on the output state. As introduced in \cite{Rosanna}, our simulation of the quantum optics experiments utilises a number of techniques to increase its efficiency. The result is a powerful simulation that allows us to simulate experiments with a truncation of up to $170$ photons (in two modes) in the order of seconds, thus allowing a broad range of exotic states -- with important contributions at high photon numbers -- to be assessed.

\subsubsection{Our three-stage algorithm}
\label{sec:three_stage}

One significant challenge in our approach is that to simulate each experiment accurately we need to truncate the Hilbert space at a high number, but the higher the truncation is, the slower the algorithm will run. Our search space of quantum experiments is huge, and to do an effective search we need to evaluate a large number of different experiments, but this becomes increasingly more computationally expensive for larger truncations. To overcome this, our GA, introduced in \cite{Rosanna}, has three stages: \\

\noindent 1.	A large number of random genomes are created and evaluated. In this stage the truncation of the Hilbert space is small (we vary this, but it's often around $30$), and therefore the quantum simulation is only approximate. But it is fast. A collection of genomes with the best fitness values are selected for the next stage. While the simulation here is only approximate, it still provides a valuable guide as to where the GA is likely to find experiments with higher fitness values -- this stage seeds the GA in the subsequent stage with a substantially better starting population than picking purely at random.  \\

\noindent 2.	We run Matlab's inbuilt GA \cite{matlabga} with a medium-sized population. The simulation is less approximate than stage 1 (because the truncation is larger, around $80$), and hence slower. This stage only runs for a set number of generations, usually 10. This stage performs a medium-speed global search, and provides the final stage with a strong population.\\

\noindent 3.	In the final stage, the simulation is accurate but slow. In this stage, the fitness function will first simulate the circuit specified by the input genome at a very low truncation, then repeat this, increasing the truncation on each iteration, until either the average number of photons in the final state converges, or until the maximum truncation is reached (where the maximum truncation is specified by the user). This ensures the results are reliable and accurate, while still running in a reasonable time. Here we again use Matlab's inbuilt GA \cite{matlabga}, but the population is smaller, and the search is more local. \\

\subsubsection{Overcoming challenges in evaluating our fitness function}
\label{sec:overcoming_challenges}

As our task is to find specific states, the obvious fitness function here is to evaluate the fidelity between the target state and the state outputted in each simulation. For example, if we wish to find a cat state, then we can calculate the fidelity to a cat state. But \emph{which} cat state should we be evaluating against? An (unnormalised) cat state given by $\ket{\alpha} + e^{i\theta}\ket{-\alpha}$ has three real-valued parameters (the value of $\theta$ and the magnitude and phase of $\alpha$), so we should compare against every combination of these parameters. Even restricting to small cat states, e.g. $|\alpha| \leq 2$, and discretising the parameters, we still have a large number of cat states to compare against (in the order of $10^5$ for this paper). This is not a problem in stages $2$ and $3$ of our algorithm, because the run-time is dominated by simulating each experiment. But in stage $1$, where the truncation is small, the overhead from evaluating the fidelity becomes significant.

This problem is exacerbated when we consider how our algorithm will commonly be used to design new quantum experiments. Generally, we expect a user to specify which states, operators, and measurements they have available, and then to run the algorithm to find which states, and to what fidelity, can be produced with their given equipment. In this case, in stage $1$ of the algorithm ideally we would search for \emph{all} of the quantum states of interest (e.g. the $5$ classes used in this paper) simultaneously, but this would require a vast number of fidelities to be computed for each simulation (around $10^6$ for this paper).

We overcome this problem by using a deep neural network (DNN) to classify each quantum state, thus bypassing the need to calculate each fidelity. The DNN is introduced in detail in the next section, but it suffices for now to consider the DNN as a black box for which we input a quantum state, and the DNN outputs a classification. More specifically, we input a quantum state into the DNN, and the DNN will output a probability distribution as to whether the state is a cat state, squeezed cat state, zombie cat state, ON state, cubic phase state, or none of the above. For example, if we input the state $\ket{\alpha=1} + \ket{\alpha=-1}$ into the DNN, it should give the output ``cat state''. We add a class of state that we call ``other'', so that we know whenever a simulation produces a state that is not close to any of our desired states. See the next section for more details of the DNN, including how we train it to perform the classification.

\begin{figure}[]
	\centering
	\includegraphics[trim=5cm 0cm 5cm 2cm, clip=true, width=0.9\linewidth]{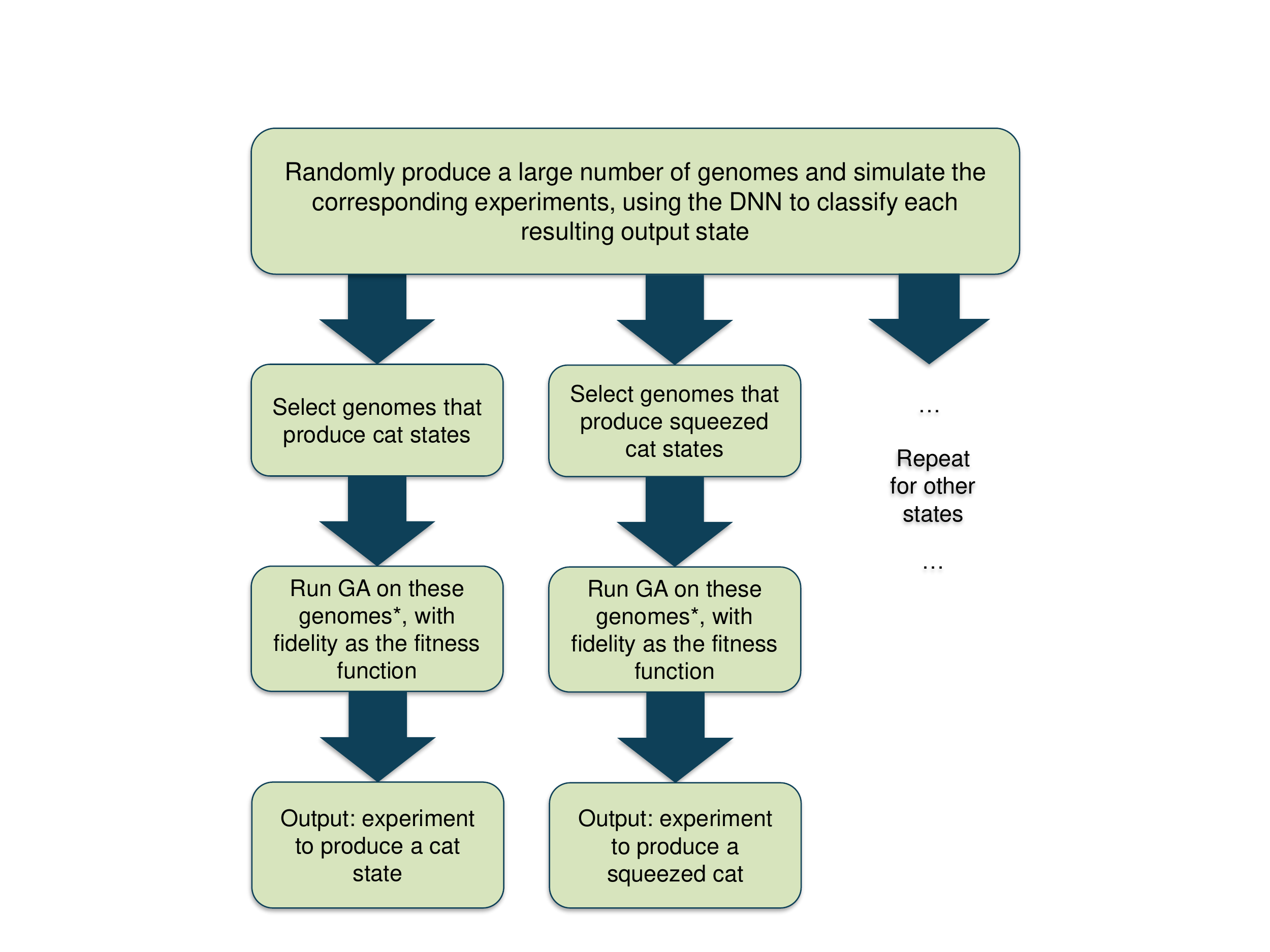}
	\caption{A flowchart of the overall structure of our hybrid algorithm that incorporates both a deep neural network (DNN) and a genetic algorithm (GA) to design experiments to produce a range of quantum states. \\ \smaller *The GA runs in stages 2 and 3 of the 3-stage algorithm discussed in Section \ref{sec:three_stage} of the main text.}
	\label{fig:Hybrid}
\end{figure}

When used in stage $1$ of our algorithm, this method of using a DNN has two significant advantages over the alternative method of calculating the fidelities. Firstly, once the DNN is trained to classify states it runs substantially faster than calculating the fidelity. Secondly, the DNN classifies for all $5$ states (or more) simultaneously. We therefore use the DNN as follows: In stage $1$, we evaluate a large number of genomes, using the DNN to classify each. We then take a number of the genomes that come closest to producing cat states, according to the DNN, and then send these genomes through stages $2$ and $3$, using the fidelity as the fitness function (population sizes for the runs of AdaQuantum in this paper are given in Appendix \ref{sec:running_Ada}). We then repeat stages $2$ and $3$ for the other $4$ states. When used in this way in stage $1$, the DNN is around two orders of magnitude faster than calculating all the fidelities. The overall structure of our hybrid algorithm that incorporates both the DNN and the GA is shown in Fig.~\ref{fig:Hybrid}.\

Our algorithm AdaQuantum is available free-to-use on GitHub \cite{AdaQ}.

\subsection{A neural network for classifying quantum states}

Having explained \emph{how} and \emph{why} we choose to utilise a DNN for classifying quantum states, we will now introduce neural networks, and explain how we construct and train ours.

\begin{figure*}[]
	\centering
	\includegraphics[width=0.7\linewidth]{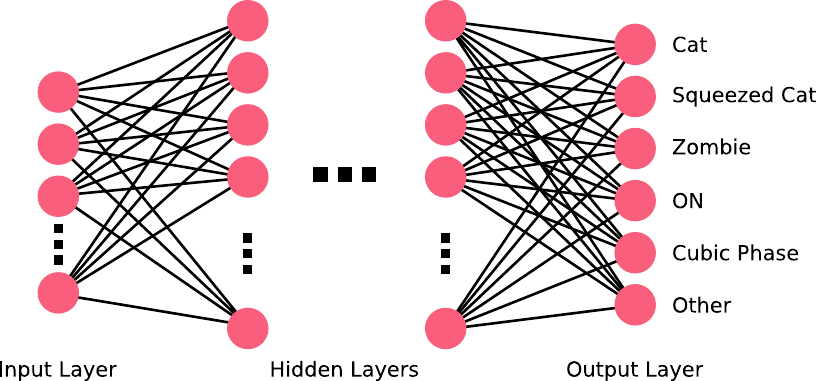}
	\caption{Our deep neural network (DNN) for classifying quantum states. We input the number distribution of a quantum state, and the DNN outputs a probability distribution corresponding to the probabilities that the inputted state was one of a fixed set of classes (introduced in Section \ref{eng_states}): cat state, squeezed cat state, zombie cat state, ON state, cubic phase state, or none of the above.}
	\label{fig:NN}
\end{figure*}

A classifier deep neural network (DNN) is a machine learning technique used to classify data into a set of classes. In our case, we input a quantum state, and the DNN will output a probability distribution as to whether the state is a cat state, squeezed cat state, zombie cat state, ON state, cubic phase state, or none of the above. Data is input to the DNN as a vector $\vec{x} \in \mathbb{R}^n$. The network is built up of layers of ``neurons''. In a given layer, each neuron is assigned a value that is calculated by first taking a linear combination of the values in the previous layer, then doing a non-linear activation function. Mathematically, this is expressed as:
\[
    \vec{x}_{i+1} = \sigma \left( M_i \vec{x_i} + \vec{b_i} \right)
\]
Where $\vec{x}_{i+1}$ are the neuron values for the next layer, $M_i$ is matrix of \emph{weights}, $\vec{b}_i \in \mathbb{R}^n$ is a \emph{bias} vector, and $\sigma$ is the activation function, which is applied element-wise.

Our final layer has 6 neurons, each corresponding to a class of state; this layer uses a different activation function than the rest of the network and produces a probability distribution, corresponding to the probability that the network thinks the input state was each of the classes (e.g. if the first neuron in the output layer is $0.9$, the network has determined that the input state has a $90\%$ probability of being a cat state). The neuron with the maximum probability therefore indicates which class the input state is predicted to belong to. The values of $M_i$ and $\vec{b}_i$ must be \emph{learnt} by the network so that it produces the desired output -- this is achieved by training, which is described below. A visualisation of the neural network is shown in Fig.~\ref{fig:NN}. \\

To train the neural network, we first need some labelled training data, which should take the form of a set of quantum states, and a label for each state (the label gives the ``class'' of each state, for example a ``cat state''). This was generated by sampling parameter values for the states from a random distribution. In other words, we first create a cat state with ramdomly generated parameters (where the parameters for the cat state are $\alpha$ and $\theta$). By associating with this state the label ``cat state'', we now have our first piece of training data. This is repeated for the cat state a number of times (approx. 1700 in this paper), before moving on to the other classes of quantum states. We repeat this process to create the testing data. Once the training (and testing) data has been generated, this can be used to train (and test) the DNN. The goal of the DNN is therefore to take in a given state, and correctly tell us which of the classifications this state belongs to.

When producing quatum states for the training and testing data, the coefficients of the quantum states in the Fock basis were then calculated using either the analytic expressions (where possible), or by matrix representations of operators acting on the vacuum state (in both cases using a truncated Hilbert space) \cite{barnett2002methods}. Some states with random coefficients were also generated (labelled \emph{other}). The generated training states are inputted to the neural network and a loss function (the softmax cross entropy) is computed using the values of the output layer, and the actual label of the state. The aim of training the network is to modify the values of the biases and weights so that this loss function is minimised; to achieve this, we used an optimisation algorithm called Adam \cite{Kingma2014Adam2}, a variation on the stochastic gradient descent method commonly used in machine learning.

Data must be inputted to the neural network as a real, finite-dimensional vector, but our quantum states belong to a complex, infinite-dimensional Hilbert space. We choose to convert the complex coefficients of each state (in the truncated Fock basis) to real numbers by taking the modulus of the Fock coefficients (throughout the paper we refer to the set of moduli of the Fock coefficients as the ``number distribution''). We tried inputting the phase information as extra inputs to the DNN, but this didn't improve the accuracy. 

A common problem encountered in machine learning is overfitting, which occurs when the model learnt by the neural network fits to outliers in your training data set, but doesn't generalise well (as an extreme example, if the DNN has enough free parameters it can ``memorise'' the whole training set given enough training iterations). One method to detect this is to split your data set in two: training and testing. The network is trained using only the data in the training set. Its accuracy can then be evaluated on the testing data set (which the network has never seen before). If the accuracy of the network on the testing set is significantly less than its accuracy on the training set, then it's likely that overfitting has occurred. Using larger training/testing data sets can help to avoid overfitting, as well as techniques such as dropout and regularisation, all of which we experimented with here.

We implemented our DNN in TensorFlow \cite{abadi2016tensorflow}. We used a training data-set containing $10,000$ states, and a testing data-set of $3,000$ states, where approximately $1/6^{\text{th}}$ of the states in each data-set belong to each class. After some experimentation, we settled on a DNN consisting of $3$ fully-connected hidden layers, comprising $25$, $25$ and $10$ neurons, respectively. After $5,000$ steps (\emph{epochs}) of training, the network classified the test data with an accuracy of $99.3\%$. Our DNN is available on GitHub \cite{DNN_git} (which includes the code for generating states for the training data). 

A standard metric to monitor how well a neural network is classifying data, aside from the accuracy, is the confusion matrix. This is a 6x6 matrix (in our case), where the rows represent the actual classes of the states, and the columns represent the classes predicted by the network. The entry in the $i^{\text{th}}$ row and $j^{\text{th}}$ column shows how many states of class $i$ were classified as class $j$. The confusion matrix calculated using our test data was:
\[
\begin{blockarray}{ccccccc}
 & \text{Cat} & \text{Sq.-Cat} & \text{Zombie} & \text{ON} & \text{CP} & \text{Other} \\
\begin{block}{c(cccccc)}
    \text{Cat} ~~& 497 & 3 & 0 & 0 & 0 & 0 \\
    \text{Sq.-Cat} ~~& 0 & 500 & 0 & 0 & 0 & 0 \\
    \text{Zombie} ~~& 1 & 0 & 499 & 0 & 0 & 0 \\
    \text{ON} & 0 ~~& 0 & 1 & 499 & 0 & 0 \\
    \text{CP} & 5 ~~& 7 & 3 & 0 & 485 & 0 \\
    \text{Other} ~~& 0 & 0 & 0 & 1 & 0 & 499 \\
\end{block}
\end{blockarray}
\]
From this we can see that the network is best at classifying squeezed cat states, however it has also classified $10$ states incorrectly as squeezed cat states. It is also clear that the network is worst at classifying cubic phase states, which is to be expected as they have the most complex structure of all the states we're interested in. Also note that no states were incorrectly classified as ``other'', meaning we were unlikely to falsely classify states that we're looking for as being useless. \\

\section{Results}





\begin{table*} [t]
{\renewcommand{\arraystretch}{1.5} 
\begin{tabular}{lllll}
		{\bf Target state} & {\bf Target parameters} & {\bf Experiment} & {\bf Experiment parameters} & {\bf Fidelity} \\
		\hline
        Cat & $\alpha = -2-i$, $\theta = 0$ & $\bra{6} \hat{U}_{T_1} \ket{z_1, z_2}$ & $z_1 = 0.701 e^{4.10i}, z_2 = 0.156 e^{0.847i}, T_1 = 0.407$ & 99.85\% \\
        Squeezed Cat & $\alpha = -0.\dot{2}+0.\dot{2}i$, $z = 1.09+0.47i$ & $\bra{4}\hat{U}_{T_2}\ket{z_3}_{12}$ & $z_3 = 1.28 e^{0.422i}, T_2 = 0.499$ & 99.78\% \\
        Zombie & $\alpha = -0.28 + 0.53i$ & $\bra{4}\hat{D}_{1}(\alpha_1)\hat{U}_{T_3}\ket{z_4, 0}$ & $z_4 = 1.26 e^{2.64i}, T_3 = 0.724, \alpha_1 = 2.16 e^{0.265i}$ & 96.84\% \\
        ON & $n = 2$, $\delta = 0.32$ & $ \bra{8}\hat{U}_{T_4}\ket{z_5}_{12}$ & $z_5 = 0.985 e^{6.28i}, T_4 = 0.606$ & 97.77\% \\
        Cubic Phase & $\gamma = 0.05, z = 0.29$ &  $\bra{5}\hat{U}_{T_5}\ket{z_6, 1}$ & $z_6 = 0.586 e^{3.14i}, T_5 = 0.612$ & 96.11\% \\
	\end{tabular}}
\caption{The results found by running AdaQuantum to find experimental schemes to produce $5$ different quantum states. Normalisation constants are omitted. Also, in the heralding measurements we omit the identity operator that acts on the remaining mode, i.e. $\bra{n} \equiv \bra{n} \otimes \mathbb{I}$, where $\bra{n}$ acts on the first mode and $\mathbb{I}$ acts on the second mode. The final column shows the fidelity of the output state with the target state. All of these states can be made using present day experimental equipment, and all experimental designs produced states with a fidelity of at least $96\%$ with their target state. See Fig.~\ref{fig:NO} for a schematic of the experiment to produce an ON state.}
\label{table_results}
\end{table*}

\begin{figure}[]
	\centering
	\includegraphics[trim=7cm 6.25cm 1cm 1cm, clip=true, width=1\linewidth]{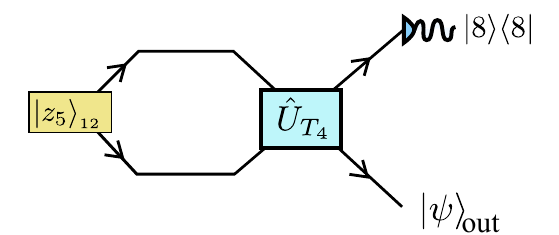}
	\caption{A sample experiment designed by AdaQuantum: this arrangement of quantum optics elements creates an ON state to a fidelity of $97.77\%$. See Table \ref{table_results} for details and parameter values. As desribed in the main text, this state is created by sending a two-mode squeezed vacuum state ($\ket{z_{5}}_{12}$) through a beam splitter ($U_{T_4}$), and then acting with the POVM $\ket{8}\bra{8}$ on the first mode of this two-mode state. }
	\label{fig:NO}
\end{figure}

We ran AdaQuantum to find the $5$ states introduced in Section \ref{eng_states}, and obtained the results in Table \ref{table_results}. The hyperparameters and settings of the genetic algorithm that were used to obtain these results are given in Appendix \ref{sec:running_Ada}. All of the states were found to a fidelity above $96\%$, and two of them were over $99.7\%$.

As an example, to produce an ON state to a fidelity of $97.77\%$ we first send a two-mode squeezed vacuum state through a beam splitter. This creates the two-mode state $\hat{U}_{T_4}\ket{z_5}_{12}$. Next, we do an $8$-photon heralding measurement on the first mode of the output. As explained in Appendix \ref{sec:apx_toolbox}, we can calculate the output state from such a heralding measurement by acting with  $\bra{8} \otimes \mathbb{I}$, where $\mathbb{I}$ is the identity (acting on the second mode). Finally, we normalise this state. The full ON state in this example is therefore given by
\begin{equation}
\label{ON_state_equ}
\mathcal{N} \left( \bra{8} \otimes \mathbb{I} \right) \hat{U}_{T_4}\ket{z_5}_{12},
\end{equation}
where $\mathcal{N}$ is the normalisation constant. Note that in Table \ref{table_results}, the normalisation constant and the identity are omitted for all states. The single-mode state given in Eq.~(\ref{ON_state_equ}) has a $97.77\%$ fidelity to the target ON state. A schematic of this experiment is shown in Fig.~\ref{fig:NO}.

Similar schemes -- with similar simplicity -- were found for all $5$ states. As discussed above, all of the experimental elements used here are accessible with current technology, with perhaps the biggest challenge in producing the states in table \ref{table_results} being the larger-number heralding measurements \cite{humphreys2015tomography,gerrits2010generation}. The purpose of this paper is to introduce the method of using such an algorithm, and demonstrate its effectiveness; future work will undertake a deeper analysis of the capabilities of AdaQuantum and the experiments and states that it has produced.

\section{Conclusion}

%
%

We have introduced a hybrid machine learning algorithm for designing quantum optics experiments. A genetic algorithm was used to search for optimal arrangements of experimental elements that produce a range of useful and interesting optical quantum states; and a deep neural network was used to speed up the evaluations of each experimental arrangement by quickly and accurately classifying quantum states. Combining these techniques, our algorithm found experimental arrangements to produce all $5$ states we asked it to, all to a high fidelity. This demonstrates the power and flexibility of the technique of using methods from artificial intelligence and machine learning to design and optimise quantum physics experiments. \\

\noindent \emph{Acknowledgements:} We thank Joseph Namara Hollis for the artwork in Fig~\ref{fig:Ada}. We acknowledge discussions with Gerardo Adesso, Ryuji Takagi, Tom Bromley and Ender {\"O}zcan. P.K. acknowledges support from the Royal Commission for the Exhibition of 1851. L.O'D. was supported by a vacation bursary from EPSRC. 

\subsection{References}


\bibliographystyle{ieeetr}
\bibliography{Rosanna_references_plus2}

\begin{thebibliography}{10}

\bibitem{dunjko2017machine}
V.~Dunjko and H.~J. Briegel, ``Machine learning and artificial intelligence in
  the quantum domain,'' {\em arXiv preprint arXiv:1709.02779}, 2017.

\bibitem{biamonte2017quantum}
J.~Biamonte, P.~Wittek, N.~Pancotti, P.~Rebentrost, N.~Wiebe, and S.~Lloyd,
  ``Quantum machine learning,'' {\em Nature}, vol.~549, no.~7671, p.~195, 2017.

\bibitem{schuld2015introduction}
M.~Schuld, I.~Sinayskiy, and F.~Petruccione, ``An introduction to quantum
  machine learning,'' {\em Contemporary Physics}, vol.~56, no.~2, pp.~172--185,
  2015.

\bibitem{knott2016search}
P.~A. Knott, ``A search algorithm for quantum state engineering and
  metrology,'' {\em New Journal of Physics}, vol.~18, no.~7, p.~073033, 2016.

\bibitem{krenn2016automated}
M.~Krenn, M.~Malik, R.~Fickler, R.~Lapkiewicz, and A.~Zeilinger, ``Automated
  search for new quantum experiments,'' {\em Physical review letters},
  vol.~116, no.~9, p.~090405, 2016.

\bibitem{melnikov2018active}
A.~A. Melnikov, H.~P. Nautrup, M.~Krenn, V.~Dunjko, M.~Tiersch, A.~Zeilinger,
  and H.~J. Briegel, ``Active learning machine learns to create new quantum
  experiments,'' {\em Proceedings of the National Academy of Sciences},
  p.~201714936, 2018.

\bibitem{arrazola2018machine}
J.~M. Arrazola, T.~R. Bromley, J.~Izaac, C.~R. Myers, K.~Br{\'{a}}dler, and
  N.~Killoran, ``Machine learning method for state preparation and gate
  synthesis on photonic quantum computers,'' {\em Quantum Science and
  Technology}, vol.~4, p.~024004, jan 2019.

\bibitem{sabapathy2018near}
K.~K. Sabapathy, H.~Qi, J.~Izaac, and C.~Weedbrook, ``Near-deterministic
  production of universal quantum photonic gates enhanced by machine
  learning,'' {\em arXiv preprint arXiv:1809.04680}, 2018.

\bibitem{ourjoumtsev2007generation}
A.~Ourjoumtsev, H.~Jeong, R.~Tualle-Brouri, and P.~Grangier, ``Generation of
  optical ‘schr{\"o}dinger cats’ from photon number states,'' {\em Nature},
  vol.~448, no.~7155, p.~784, 2007.

\bibitem{huang2015optical}
K.~Huang, H.~Le~Jeannic, J.~Ruaudel, V.~B. Verma, M.~D. Shaw, F.~Marsili, S.~W.
  Nam, E.~Wu, H.~Zeng, Y.-C. Jeong, {\em et~al.}, ``Optical synthesis of
  large-amplitude squeezed coherent-state superpositions with minimal
  resources,'' {\em Physical review letters}, vol.~115, no.~2, p.~023602, 2015.

\bibitem{etesse2015experimental}
J.~Etesse, M.~Bouillard, B.~Kanseri, and R.~Tualle-Brouri, ``Experimental
  generation of squeezed cat states with an operation allowing iterative
  growth,'' {\em Physical review letters}, vol.~114, no.~19, p.~193602, 2015.

\bibitem{gottesman2001encoding}
D.~Gottesman, A.~Kitaev, and J.~Preskill, ``Encoding a qubit in an
  oscillator,'' {\em Physical Review A}, vol.~64, no.~1, p.~012310, 2001.

\bibitem{Note1}
The algorithm, AdaQuantum, is named after Ada Lovelace, the worlds first
  computer programmer, and resident of Nottingham, where our own algorithm was
  born.

\bibitem{Rosanna}
R.~Nichols, L.~Mineh, J.~Rubio, J.~C.~F. Matthews, and P.~A. Knott, ``Designing
  quantum experiments with a genetic algorithm,'' {\em arXiv preprint
  arXiv:1812.01032}, 2018.

\bibitem{jin2011surrogate}
Y.~Jin, ``Surrogate-assisted evolutionary computation: Recent advances and
  future challenges,'' {\em Swarm and Evolutionary Computation}, vol.~1, no.~2,
  pp.~61--70, 2011.

\bibitem{rubio2018non}
J.~Rubio, P.~Knott, and J.~Dunningham, ``Non-asymptotic analysis of quantum
  metrology protocols beyond the {C}ram{\'e}r--rao bound,'' {\em Journal of
  Physics Communications}, vol.~2, no.~1, p.~015027, 2018.

\bibitem{bartley2012multiphoton}
T.~J. Bartley, G.~Donati, J.~B. Spring, X.-M. Jin, M.~Barbieri, A.~Datta, B.~J.
  Smith, and I.~A. Walmsley, ``Multiphoton state engineering by heralded
  interference between single photons and coherent states,'' {\em Physical
  Review A}, vol.~86, no.~4, p.~043820, 2012.

\bibitem{gerrits2010generation}
T.~Gerrits, S.~Glancy, T.~S. Clement, B.~Calkins, A.~E. Lita, A.~J. Miller,
  A.~L. Migdall, S.~W. Nam, R.~P. Mirin, and E.~Knill, ``Generation of optical
  coherent-state superpositions by number-resolved photon subtraction from the
  squeezed vacuum,'' {\em Physical Review A}, vol.~82, no.~3, p.~031802, 2010.

\bibitem{mehmet2011squeezed}
M.~Mehmet, S.~Ast, T.~Eberle, S.~Steinlechner, H.~Vahlbruch, and R.~Schnabel,
  ``Squeezed light at 1550 nm with a quantum noise reduction of 12.3 db,'' {\em
  Opt. Express}, vol.~19, no.~25, pp.~25763--25772, 2011.

\bibitem{muller2015coherent}
K.~M{\"u}ller, A.~Rundquist, K.~A. Fischer, T.~Sarmiento, K.~G. Lagoudakis,
  Y.~A. Kelaita, C.~{Sanchez Mu{\~n}oz}, E.~del Valle, F.~P. Laussy, and
  J.~Vu{\v{c}}kovi{\'c}, ``Coherent generation of nonclassical light on chip
  via detuned photon blockade,'' {\em Phys. Rev. Lett.}, vol.~114, no.~23,
  p.~233601, 2015.

\bibitem{claudon2010highly}
J.~Claudon, J.~Bleuse, N.~S. Malik, M.~Bazin, P.~Jaffrennou, N.~Gregersen,
  C.~Sauvan, P.~Lalanne, and J.~G{\'e}rard, ``A highly efficient single-photon
  source based on a quantum dot in a photonic nanowire,'' {\em Nature Photon.},
  vol.~4, no.~3, pp.~174--177, 2010.

\bibitem{morin2012high}
O.~Morin, V.~D’Auria, C.~Fabre, and J.~Laurat, ``High-fidelity single-photon
  source based on a type {II} optical parametric oscillator,'' {\em Opt.
  Lett.}, vol.~37, no.~17, pp.~3738--3740, 2012.

\bibitem{ourjoumtsev2006quantum}
A.~Ourjoumtsev, R.~Tualle-Brouri, and P.~Grangier, ``Quantum homodyne
  tomography of a two-photon {F}ock state,'' {\em Phys. Rev. Lett.}, vol.~96,
  no.~21, p.~213601, 2006.

\bibitem{huang2015thesis}
K.~Huang, {\em Optical Hybrid Architectures for Quantum Information
  Processing}.
\newblock PhD thesis, l'{\'E}cole Normale Sup{\'e}rieure de Paris and East
  China Normal University., 2015.

\bibitem{humphreys2015tomography}
P.~C. Humphreys, B.~J. Metcalf, T.~Gerrits, T.~Hiemstra, A.~E. Lita, J.~Nunn,
  S.~W. Nam, A.~Datta, W.~S. Kolthammer, and I.~A. Walmsley, ``Tomography of
  photon-number resolving continuous-output detectors,'' {\em arXiv preprint
  arXiv:1502.07649}, 2015.

\bibitem{lee2015quantum}
S.-Y. Lee, C.-W. Lee, J.~Lee, and H.~Nha, ``Quantum phase estimation using a
  class of entangled states: {NOON}-type states,'' {\em arXiv preprint
  arXiv:1505.06000}, 2015.

\bibitem{barnett2002methods}
S.~M. Barnett and P.~M. Radmore, {\em Methods in theoretical quantum optics},
  vol.~15.
\newblock Oxford University Press, 2002.

\bibitem{schrodinger1935gegenwartige}
E.~Schr{\"o}dinger, ``Die gegenw{\"a}rtige situation in der quantenmechanik,''
  {\em Naturwissenschaften}, vol.~23, no.~49, pp.~823--828, 1935.

\bibitem{knott2016practical}
P.~Knott, T.~Proctor, A.~Hayes, J.~Cooling, and J.~Dunningham, ``Practical
  quantum metrology with large precision gains in the low-photon-number
  regime,'' {\em Physical Review A}, vol.~93, no.~3, p.~033859, 2016.

\bibitem{lee2015quantum_cat}
S.-Y. Lee, C.-W. Lee, H.~Nha, and D.~Kaszlikowski, ``Quantum phase estimation
  using a multi-headed cat state,'' {\em JOSA B}, vol.~32, no.~6,
  pp.~1186--1192, 2015.

\bibitem{jiang2016dynamics}
L.-y. Jiang, Q.~Guo, X.-x. Xu, M.~Cai, W.~Yuan, and Z.-l. Duan, ``Dynamics and
  nonclassical properties of an opto-mechanical system prepared in four-headed
  cat state and number state,'' {\em Optics Communications}, vol.~369,
  pp.~179--188, 2016.

\bibitem{rivas2012sub}
A.~Rivas and A.~Luis, ``Sub-heisenberg estimation of non-random phase shifts,''
  {\em New Journal of Physics}, vol.~14, no.~9, p.~093052, 2012.

\bibitem{hall2012heisenberg}
M.~J.~W. Hall and H.~M. Wiseman, ``Heisenberg-style bounds for arbitrary
  estimates of shift parameters including prior information,'' {\em New J.
  Phys.}, vol.~14, no.~3, p.~033040, 2012.

\bibitem{hall2012universality}
M.~J.~W. Hall, D.~W. Berry, M.~Zwierz, and H.~M. Wiseman, ``Universality of the
  {H}eisenberg limit for estimates of random phase shifts,'' {\em Phys. Rev.
  A}, vol.~85, no.~4, p.~041802, 2012.

\bibitem{sabapathy2018states}
K.~K. Sabapathy and C.~Weedbrook, ``On states as resource units for universal
  quantum computation with photonic architectures,'' {\em Phys. Rev. A},
  vol.~97, no.~6, p.~062315, 2018.

\bibitem{kok2010introduction}
P.~Kok and B.~W. Lovett, {\em Introduction to optical quantum information
  processing}.
\newblock Cambridge University Press, 2010.

\bibitem{ghose2007non}
S.~Ghose and B.~C. Sanders, ``Non-gaussian ancilla states for continuous
  variable quantum computation via gaussian maps,'' {\em Journal of Modern
  Optics}, vol.~54, no.~6, pp.~855--869, 2007.

\bibitem{takagi2018convex}
R.~Takagi and Q.~Zhuang, ``Convex resource theory of non-gaussianity,'' {\em
  Phys. Rev. A}, vol.~97, p.~062337, Jun 2018.

\bibitem{gu2009quantum}
M.~Gu, C.~Weedbrook, N.~C. Menicucci, T.~C. Ralph, and P.~van Loock, ``Quantum
  computing with continuous-variable clusters,'' {\em Physical Review A},
  vol.~79, no.~6, p.~062318, 2009.

\bibitem{matlabga}
T.~M. Inc., ``Global optimization toolbox: User's guide (r2018a).''
\newblock \url{https://uk.mathworks.com/help/gads/index.html}. Last accessed
  2018-07-17.

\bibitem{AdaQ}
``{AdaQuantum} is available and free to use on {GitHub}.''
  https://github.com/paulk444/AdaQuantum, 2019.

\bibitem{Kingma2014Adam2}
D.~P. Kingma and J.~Ba, ``{Adam: A Method for Stochastic Optimization}.''
  arXiv:1412.6980 [cs.LG], 12 2014.

\bibitem{abadi2016tensorflow}
M.~Abadi, P.~Barham, J.~Chen, Z.~Chen, A.~Davis, J.~Dean, M.~Devin,
  S.~Ghemawat, G.~Irving, M.~Isard, {\em et~al.}, ``Tensorflow: a system for
  large-scale machine learning.,'' in {\em OSDI}, vol.~16, pp.~265--283, 2016.

\bibitem{DNN_git}
``Our {DNN} and quantum-state generator on {GitHub}.''
  https://github.com/lewis-od/Quantum-Optics, 2018.

\bibitem{paris1996displacement}
M.~G.~A. Paris, ``Displacement operator by beam splitter,'' {\em Phys. Lett.
  A}, vol.~217, no.~2, pp.~78--80, 1996.

\bibitem{nielsen2010quantum}
M.~A. Nielsen and I.~L. Chuang, {\em Quantum Computation and Quantum
  Information}.
\newblock Cambridge University Press, 2010.

\bibitem{deep2007new}
K.~Deep and M.~Thakur, ``A new mutation operator for real coded genetic
  algorithms,'' {\em Applied mathematics and Computation}, vol.~193, no.~1,
  pp.~211--230, 2007.

\end{thebibliography}

\begin{widetext}
\appendix

\section{Quantum optics toolbox details}
\label{sec:apx_toolbox}

\noindent\emph{Input states -} The squeezed vacuum is given by $|z\rangle = \hat{S}(z) |0\rangle$ where the squeezing operator is $\hat{S}(z)=\exp{ \left[ {1 \over 2} (z^* {\hat{a}}^{^2} - z {\hat{a}^{{\dagger}^2}}) \right] }$ and $z=r e^{i \theta_s}$ where $r$ is the (positive and real) amplitude, $\theta_s \in [0,2\pi]$ is the squeezing angle and $\hat{a}$ ($\hat{a}^{\dagger}$) is the annihilation (creation) operator. Squeezed states can be made up to $r \approx 1.4$, but this is extremely challenging experimentally so we set the limit to $r=1.3$ \cite{mehmet2011squeezed}. Similarly, the two mode squeezed vacuum is given by $|z\rangle_{12} = \hat{S}_{12}(z) |0,0\rangle$, where the two mode squeezing operator is $\hat{S}_{12}(z)=\exp{ (z^* \hat{a}\hat{b} - z \hat{a}^{\dagger}\hat{b}^{\dagger}) }$, where $\hat{a}$ and $\hat{b}$ act on modes $1$ and $2$, respectively, and again $z$ is complex. The coherent state is given by $|\alpha\rangle = \hat{D}(\alpha) |0\rangle$ where the displacement operator is $\hat{D}(\alpha) = \exp{ (\alpha \hat{a}^{\dagger} - \alpha^* \hat{a}) }$, $\alpha = |\alpha| e^{i \theta_c}$ where $|\alpha|$ is the  amplitude, and $\theta_c \in [0,2\pi]$ is the coherent state phase. The amplitude of the coherent state can be large in experiments, so instead it is limited by the numerical methods we use: we set the limit to $\alpha=4$. The final input state is the Fock state of which the simplest is the vacuum $|0\rangle$. Single photons, $|1\rangle$, can be emitted from a quantum dot \cite{muller2015coherent,claudon2010highly} or heralded \cite{morin2012high}. We also consider the two photon state, $|2\rangle$, which has been made in \cite{ourjoumtsev2006quantum,huang2015thesis}. Higher number Fock states can be made, e.g. by heralding, but are challenging to produce to a high fidelity and are not included here. \\

\noindent\emph{Operators -} The beam splitter is described by the unitary operator $\hat{U}_{T}=e^{-i\theta_b(e^{i\phi_b}\hat{a}^{\dagger}\hat{b}+e^{-i\phi_b}\hat{a}\hat{b}^{\dagger})}$, where $\hat{a}$ and $\hat{b}$ are annihilation operators for the two modes, and we choose the arbitrary phase to be $\phi_b=-\pi/2$. Here $T=\cos^2{\theta_b}$ is the transmissivity of the beam splitter and therefore for a 50:50 beam splitter $\theta_b=\pi/4$ giving $\hat{U}_{T=50}$. Next, the displacement operator, $\hat{D}(\beta)$ (defined above), is implemented by mixing the state with a large local oscillator at a highly transmissive beam splitter \cite{paris1996displacement} ($\beta$ has the same restrictions as $\alpha$). The phase operator is given by $e^{i\hat{n}\theta}$ where $\hat{n}=\hat{a}^{\dagger}\hat{a}$ and $\theta \in [0,2\pi]$. \\ 

\noindent\emph{Measurements -} After we have applied a number of operators we perform a heralding measurement on one mode of the final state. For example, if we wish to herald on the one photon state we can perform a number resolving detection \cite{humphreys2015tomography,gerrits2010generation}, and only keep runs that measure one photon. The measurement is given by a projection \cite{nielsen2010quantum}: to follow the single photon example we project with $|1\rangle \langle 1| \otimes \hat{\mathbb{I}}$. We are then left with a separable state $|1\rangle \otimes |\psi_f\rangle$, but we can ignore the measurement mode, and after normalisation we are left with the final one mode state: $|\psi_f\rangle$. This whole process can be more easily modeled by acting on the two-mode, pre-measurement state with $\langle 1| \otimes \hat{\mathbb{I}}$. In the main text we drop the identity and just write $\langle 1|$, and this measurement is always performed on the first mode of Fig.~\ref{fig:Random_general_scheme}.

\section{Running AdaQuantum}
\label{sec:running_Ada}

To obtain the results in this paper, we ran AdaQuantum with the following settings (see \cite{Rosanna} for a detailed description of how AdaQuantum works): The population sizes for stages 1, 2 and 3 are $8$x$10^6$, $5$x$10^6$, and $10^4$, respectively. Stages 2 and 3 use Matlab's build-in genetic algorithm \cite{matlabga}, and both these stages have the same hyperparameters (except for the population sizes). We use the \texttt{scattered} crossover function, with crossover fraction $0.3$. We use \texttt{tournament} selection, with a tournament size of 8. Matlab didn't have a mutation function with enough flexibility for our purposes, so we introduced a mutation functions that we call \texttt{power mutation}, which is based on \cite{deep2007new}. In short, \texttt{power mutation} mutates every gene in the genome by a random distance, whose maximum magnitude is determined by the value of a hyperparameter named \texttt{power}, for which \texttt{power}$ =1$ will mutate each gene to a completely random new value, whereas \texttt{power}$ =\infty$ doesn't mutate at all. Here we used \texttt{power}$ =10$. The number of elite children was 10.

\end{widetext}

\end{document}